# Interference of Spin Splittings in Magneto-Oscillation Phenomena in Two-Dimensional Systems

S. A. Tarasenko* and N. S. Averkiev

*Ioffe Physicotechnical Institute, Russian Academy of Sciences,
Politekhnicheskaya ul. 26, St. Petersburg, 194021 Russia*
*e-mail: tarasenko@coherent.ioffe.rssi.ru
Received April 10, 2002; in final form, April 30, 2002

The spin splitting caused by the terms linear in wavevector in the effective Hamiltonian containing can give rise to the new magneto-oscillation phenomena in two-dimensional systems. It is shown that the joint action of the spin-dependent contributions due to the heterostructure asymmetry and to the lack of inversion center in the bulk material suppresses beats that arise in the magneto-oscillation phenomena in the presence of the terms of only one of these types. © 2002 MAIK "Nauka/Interperiodica".

Thermodynamic and kinetic coefficients such as heat capacity, magnetic susceptibility, conductivity, etc. oscillate in systems with a degenerate electron gas exposed to a quantizing magnetic field at low temperatures. Such a behavior of these coefficients is due to the appearance of the Landau levels, which successively intersect the Fermi level as the magnetic field increases. Measurements of the conductivity oscillations (Shubnikov–de Haas effect) and the oscillations of magnetic susceptibility (de Haas–van Alphen effect) are among the most efficient methods of structure characterization and determination of the carrier concentrations and relaxation times.

Quantum phenomena are highly sensitive to the fine structure of the carrier energy spectrum, so that even a small spin splitting may qualitatively modify the oscillation pattern. The terms linear in wavevector $\mathbf{k}$ in the effective Hamiltonian remove the degeneracy in the carrier spectrum. In a magnetic field, the spin splitting at the Fermi surface gives rise to the oscillations with close frequencies, i.e., to beats [1]. Such a behavior of the Shubnikov–de Haas effect was observed in two-dimensional systems with a hole channel at the silicon surface [2], with the electron gas in quantum wells based on narrow-band [3, 4] and wide-band [5] semiconductors, and in other structures. The zero-field spin splittings at the Fermi level were determined from the analysis of experimental data.

In the general case, the terms linear in $\mathbf{k}$ appear because the symmetry of heterostructures is lower than the symmetry of bulk materials. In the quantum wells grown on the basis of semiconductors with zinc blende lattice in the [001] orientation, there are two types of linear contributions to the effective electron Hamiltonian. First, they originate from the cubic terms in the Hamiltonian of a bulk material without inversion center. Averaging these cubic terms along the quantization axis in the case of low subband filling with carriers gives rise to the terms linear in $\mathbf{k}$ (BIA terms), where $\mathbf{k}$ is the wavevector in the electron gas plane [6]. Furthermore, a linear contribution can be caused by the intrinsic heterostructure asymmetry which is unrelated to the crystal lattice (Rashba terms) [1]. The relative intensities of these contributions to the effective Hamiltonian of a two-dimensional electron gas can change on passing from narrow-band to wide-band semiconductors [7]. Moreover, one can control the degree of heterostructure asymmetry, e.g., by applying an electric field perpendicular to the quantum well plane. Inasmuch as the physical nature and symmetry of the BIA terms are different from those of the Rashba terms, the direct addition of their contributions to the spin splitting would be incorrect.

In this work, we demonstrate that the BIA and Rashba contributions interfere in the magneto-oscillation phenomena. The presence of only one type of linear terms gives rise to the beats. However, if the intensities of both contributions are equal, the oscillations occur only at a single frequency and the beats disappear, although the Hamiltonian contains linear terms. The suppression of beats should take place in all magneto-oscillation phenomena. We take the Shubnikov–de Haas effect as an example to examine the interference of the BIA and Rashba contributions in detail. The magnetoresistance tensor will be calculated for zero temperature taking into account either only one contribution or both terms with the same intensity. The Zeeman splitting of electronic levels will be disregarded in this work, because it is small compared to the spacing between the Landau levels in the majority of semiconductor structures based on the III-V compounds in a magnetic field perpendicular to the plane of electron gas.

Qualitatively, the disappearance of beats can be understood by analyzing the electronic spectrum in a zero magnetic field. In the absence of magnetic field, the effective Hamiltonian has the form

$$\hat{H} = \frac{\hbar^2 k^2}{2m^*} + \hat{H}_{BIA} + \hat{H}_R, \quad (1)$$

where $k = |\mathbf{k}|$ and $m^*$ is the effective mass. For a quantum well grown along the [001] direction, the spin-dependent BIA and Rashba contributions to the Hamiltonian of a two-dimensional electron gas can conveniently be written in the crystal system of coordinates ($x \parallel [1\bar{1}0]$, $y \parallel [110]$, $z \parallel [001]$):

$$\begin{aligned}\hat{H}_{BIA} &= \alpha(\hat{\sigma}_x k_y + \hat{\sigma}_y k_x), \\ \hat{H}_R &= \beta(\hat{\sigma}_x k_y - \hat{\sigma}_y k_x),\end{aligned} \quad (2)$$

where $\hat{\sigma}_x$ and $\hat{\sigma}_y$ are the Pauli matrices.

In the presence of only one type of terms linear in $\mathbf{k}$, e.g., the BIA terms, the electronic spectrum is isotropic and consists of two different spin subbands (Fig. 1a):

$$E_\pm(k) = \frac{\hbar^2 k^2}{2m^*} \pm \alpha k. \quad (3)$$

If both contributions are essential, the spectrum becomes more complicated, and the energy becomes dependent on the direction of wavevector $\mathbf{k}$ [8]. However, the spectrum is simplified if the BIA and Rashba terms have the same intensity, i.e., if $|\alpha| = |\beta|$. In this case, the spectrum consists of two identical paraboloids shifted relative to each other in the $\mathbf{k}$ space. For example, if $\alpha = \beta$ (Fig. 1b), the paraboloids are shifted along $k_y$ and characterized by the spin states with $|\pm 1/2\rangle$ projections onto the $x$ axis. Accordingly, the electronic spectrum has the form

$$E_\pm(\mathbf{k}) = \frac{\hbar^2}{2m^*}[k_x^2 + (k_y \pm k_0)^2] - \frac{m^* \gamma^2}{2\hbar^2}, \quad (4)$$

where $k_0 = \gamma m^*/\hbar^2$ and $\gamma = 2\alpha = 2\beta$.

The magneto-oscillation frequencies of kinetic coefficients depend on the Fermi surface. In the presence of only one type of linear terms (Fig. 1a), the Fermi surfaces of two spin subbands are different because of the spin splitting $2\alpha k_F$, where $\hbar k_F$ is the Fermi momentum. The subbands are responsible for the oscillations with close frequencies, giving rise to beats. In the case that the intensities of the BIA and Rashba contributions are identical (Fig. 1b), the spin subbands are equivalent and the oscillations have the same frequency, and the beats do not arise.

In the regime of small Shubnikov–de Haas oscillations and electron scattering by short-range defects, the conductivity tensor can be represented as the sum of classical magnetoresistance and the oscillating contribution,

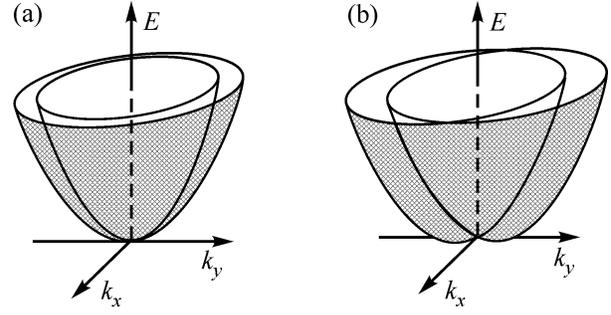

**Fig. 1.** Electronic energy spectrum in zero magnetic field for the case where (a) only one type of linear terms and (b) both contributions with equal intensities ($\alpha = \beta$) are taken into account.

$$\begin{aligned}\sigma_{xx} &= \frac{Ne^2\tau/m^*}{1+(\omega_c\tau)^2}\left\{1 + \frac{2(\omega_c\tau)^2}{1+(\omega_c\tau)^2}\delta\right\}, \\ \sigma_{xy} &= -\frac{Ne^2\omega_c\tau^2/m^*}{1+(\omega_c\tau)^2}\left\{1 - \frac{1+3(\omega_c\tau)^2}{[1+(\omega_c\tau)^2](\omega_c\tau)^2}\delta\right\},\end{aligned} \quad (5)$$

where $N$ is the two-dimensional electron concentration, $\tau$ is the momentum relaxation time, $\omega_c = eB/m^*c$ is the cyclotron frequency, $B$ is the magnetic field, $e$ is the elementary charge, and $c$ is the velocity of light. The explicit expression for the quantity $\delta$ oscillating in a magnetic field depends on the Fermi surface. In two-dimensional systems with a simple band, $\delta$ has only one harmonic [9, 10].

The calculation shows that, in the presence of only one type of terms linear in $\mathbf{k}$, the oscillating quantity has the form

$$\delta = 2\exp\left(-\frac{\pi}{\omega_c\tau}\right)\cos\left(2\pi\frac{E_F}{\hbar\omega_c}\right)\cos\left(2\pi\frac{\alpha k_F}{\hbar\omega_c}\right), \quad (6)$$

where $E_F$ is the Fermi energy measured from the subband bottom in the absence of the linear terms. The dependence of $\delta$ on magnetic field has the form of beats, because $E_F \gg \alpha k_F$.

In the case that the intensities of the BIA and Rashba terms are equal, the oscillations, as in the absence of spin splitting, contain only one harmonic,

$$\delta = -2\exp\left(-\frac{\pi}{\omega_c\tau}\right)\cos\left(2\pi\frac{E_F'}{\hbar\omega_c}\right), \quad (7)$$

where $E_F' = E_F + m^*\gamma^2/(2\hbar^2)$ is the Fermi level measured from the bottom of the subbands.

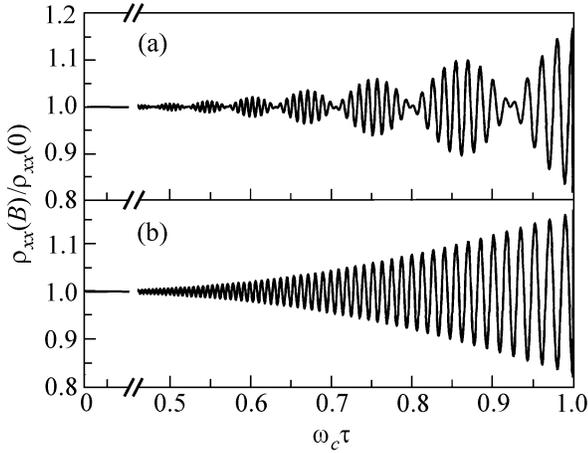

**Fig. 2.** Magnetic-field dependence of the resistivity $\rho_{xx}$ in the regime of Shubnikov–de Haas oscillations in the presence of (a) only one type of linear terms, $E_F\tau/\hbar = 50$ and $\alpha k_F/\hbar = 3$, and (b) both contributions with equal intensities and $E_F'\tau/\hbar = 50$.

Figure 2 displays the magnetic-field dependence of the resistivity

$$\rho_{xx} = \frac{\sigma_{xx}}{\sigma_{xx}^2 + \sigma_{xy}^2}.$$

If only one type of linear terms dominates (Fig. 2a), the spectrum of Shubnikov–de Haas oscillations contains two harmonics with close frequencies, and the field dependence of resistivity appears as beats. The frequency difference is determined by the spin splitting $2\alpha k_F$ at the Fermi level. In the case of identical intensities of the BIA and Rashba terms, $|\alpha| = |\beta|$ (Fig. 2b), the oscillations of both spin subbands have the same frequency and the beats are not observed.

In this work, a consistent theory of magneto-oscillation effects is developed using the Green's function method taking into account the terms linear in **k**. In the two-dimensional systems, small oscillations are observed in classical magnetic fields, $\omega_c\tau \leq 1$, while the quantity $\exp(-\pi/\omega_c\tau)$ serves as a parameter determining the oscillation amplitude [9, 10]. We assume that the inequality $E_F\tau/\hbar \gg 1$ providing good conductivity is fulfilled. In the self-consistent Born approximation, one-particle electronic Green's function for electron scattering by the short-range defects has the form

$$\hat{G}_\varepsilon(\mathbf{r}, \mathbf{r}') = \sum_{nk_y s} \frac{\Psi_{nk_y s}(\mathbf{r})\Psi_{nk_y s}^\dagger(\mathbf{r}')}{\varepsilon + E_F - E_{ns} - X_\varepsilon}, \quad (8)$$

where $\Psi_{nk_y s}(\mathbf{r})$ are the spinor electron wave functions in a magnetic field $\mathbf{B} \parallel z$ with the Landau-gauge vector potential $\mathbf{A} = (0, Bx, 0)$; $E_{ns}$ are the electronic levels; $X_\varepsilon$ is the self-energy part of the Green's function; and $n$, $k_y$, and $s = \pm$ are the quantum numbers. Green's function (8) is a $2 \times 2$ matrix in the spin space.

In the presence of only one type of linear terms, the orbital and spin states are mixed (see [11]). At $\alpha k_F \gg \hbar\omega_c$, the electron energies near the Fermi level are

$$E_{n\pm} = \hbar\omega_c(n+1) \pm \alpha\sqrt{2(n+1)}/\lambda_B. \quad (9)$$

It is precisely the splitting $2\alpha\sqrt{2(n+1)}/\lambda_B$ in the spectrum (9) which gives rise to the beats in the magneto-oscillation phenomena.

If $|\alpha| = |\beta|$, then the wave functions are the products of the spin and orbital functions, and the electronic spectrum is determined by the expression

$$E_{n\pm} = \hbar\omega_c(n+1/2) - m^*\gamma^2/2\hbar^2. \quad (10)$$

In this case, the spectrum is not split, for which reason the oscillations due to the spin subbands have the same frequency and, hence, there will be no beats.

For the scattering from the short-range potentials, the self-energy part of the Green's function does not depend on $n$ [9] and, in both cases considered, on $s$ and satisfies the equation

$$X_\varepsilon = \frac{\hbar\omega_c}{\pi}\frac{\hbar}{4\tau}\sum_{ns}\frac{1}{\varepsilon + E_F - E_{ns} - X_\varepsilon}. \quad (11)$$

In the presence of only one type of linear terms, Eq. (11) has, to first order in parameter $\exp(-\pi/\omega_c\tau)$, the solution

$$X_\varepsilon = -\frac{i\hbar}{2\tau}\Bigg[1 + 2\exp\!\left(-\frac{\pi}{\omega_c\tau}\right)$$
$$\times \exp\!\left(2\pi i\frac{\varepsilon + E_F}{\hbar\omega_c}\mathrm{sgn}\,\varepsilon\right)\cos\!\left(2\pi\frac{\alpha k_F}{\hbar\omega_c}\right)\Bigg]\mathrm{sgn}\,\varepsilon. \quad (12)$$

The multiplier $\cos(2\pi\alpha k_F/\hbar\omega_c)$ in the self-energy part (12) appears due to the level splitting. It is responsible for the beats in the magneto-oscillation phenomena, e.g., in the Shubnikov–de Haas effect.

If $|\alpha| = |\beta|$, then the self-energy part contains no such term:

$$X_\varepsilon = -\frac{i\hbar}{2\tau}\Bigg[1 - 2\exp\!\left(-\frac{\pi}{\omega_c\tau}\right)$$
$$\times \exp\!\left(2\pi i\frac{\varepsilon + E_F'}{\hbar\omega_c}\mathrm{sgn}\,\varepsilon\right)\Bigg]\mathrm{sgn}\,\varepsilon. \quad (13)$$

The Green's function allows the calculation of various kinetic and thermodynamic coefficients. Using the standard methods [12, 13], one can deduce expressions (5)–(7) for the conductivity tensor in the regime of Shubnikov–de Haas oscillations. The remaining components are related by the expressions $\sigma_{xx} = \sigma_{yy}$ and $\sigma_{yx} = -\sigma_{xy}$.

The interference of the spin-dependent contributions caused by the heterostructure asymmetry and the lack of inversion center in the bulk material was considered in [8], where the anisotropy of electronic spectrum was taken as an example. Such an anisotropy was experimentally observed in the Raman studies of the GaAs/AlGaAs structures [14]. The fact that the terms linear in **k** nonadditively add together in the case of weak localization was predicted in [15] and observed in the measurements of anomalous magnetoresistance in [16]. It was shown in [17] that the joint action of both spin-dependent contributions gives rise to the anisotropy of relaxation times in the quantum well plane. It has been demonstrated in this work that the interference of the BIA and Rashba terms linear in **k** qualitatively alters the pattern of magneto-oscillation phenomena in two-dimensional systems.

This work was supported by the Russian Foundation for Basic Research, INTAS, the program of the Presidium of the Russian Academy of Sciences "Low-Dimensional Quantum Structures," and programs of the Ministry of Industry, Science, and Technologies of the Russian Federation.

*Translated by V. Sakun*